\documentclass[fleqn,usenatbib]{mnras}

\usepackage{newtxtext,newtxmath}

\usepackage[T1]{fontenc}
\usepackage{ae,aecompl}

\usepackage{graphicx}	
\usepackage{amsmath}	
\usepackage{amssymb}	
\usepackage{booktabs}
\usepackage{array}
\usepackage{multirow}
\usepackage{float}
\usepackage{caption}

\title[CII 158$\mu$m in $z>$8 galaxies]{The Absence of  [C{\sc ii}]158$\mu$m Emission in Spectroscopically-Confirmed Galaxies at $z>8$}

\author[N. Laporte et al.]{
N. Laporte,$^{1}$\thanks{E-mail: n.laporte@ucl.ac.uk}
H. Katz,$^{2}$
R. S. Ellis,$^{1}$
G. Lagache,$^{3}$
F. E. Bauer,$^{4,5,6}$
F. Boone,$^{7}$
\newauthor 
A. K. Inoue,$^{8,9}$
T. Hashimoto,$^{9}$
H. Matsuo,$^{10}$
K. Mawatari,$^{11}$
and Y. Tamura.$^{12}$
\\
$^{1}$ Department of Physics \& Astronomy, University College London, London, WC1E 6BT, UK\\
$^{2}$ Astrophysics, University of Oxford, Denys Wilkinson Building, Keble Road, Oxford OX1 3RH, UK\\
$^{3}$ Aix-Marseille Univ., CNRS, LAM, Laboratoire d'Astrophysique de Marseille, 13388 Marseille, France \\
$^{4}$ Instituto de Astrof\'isica, Facultad de F\'isica, Pontificia Universidad Cat\'olica de Chile Av.  Vicu\~na Mackenna 4860, \\
782-0436 Macul,Santiago, Chile\\
$^{5}$ Millennium Institute of Astrophysics (MAS), Nuncio Monse\~nor S\'otero Sanz 100, Providencia, Santiago, Chile \\
$^{6}$ Space Science Institute, 4750 Walnut Street, Suite 205, Boulder, CO 80301, USA \\
$^{7}$ Institut de Recherche en Astrophysique et Plan\'etologie (IRAP), Universit\'e de Toulouse, CNRS, UPS, 31400 Toulouse, France \\
$^{8}$ Department of Physics, School of Advanced Science and Engineering, Waseda University, 3-4-1, Okubo, Shinjuku, Tokyo 169-8555, Japan \\
$^{9}$ Waseda Research Institute for Science and Engineering, 3-4-1, Okubo, Shinjuku, Tokyo 169-8555, Japan \\
$^{10}$ National Astronomical Observatory of Japan, Mitaka, Tokyo 181-8588, Japan \\
$^{11}$ Institute for Cosmic Ray Research, University of Tokyo, 5-1-5, Kashiwa-no-ha, Kashiwa, Chiba,  277-8582, Japan  \\
$^{12}$ Division of Particle and Astrophysical Science, Graduate School of Science, Nagoya University, Nagoya 464-8602, Japan.
}
\date{Accepted 2019 June 5. Received 2019 May 23; in original form 2019 April 25}
\pubyear{2019}
\begin{document}
\label{firstpage}
\pagerange{\pageref{firstpage}--\pageref{lastpage}}
\maketitle

\begin{abstract}
The scatter in the relationship between the strength of  [C{\sc ii}]158$\mu$m emission and the star formation rate at high-redshift has been the source of much recent interest. Although the relationship is well-established locally, several intensely star-forming galaxies have been found whose [C{\sc ii}]158$\mu$m emission is either weak, absent or spatially offset from the young stars. Here we present new ALMA data for the two most distant, gravitationally-lensed and spectroscopically-confirmed galaxies, A2744\_YD4 at $z=$8.38 and  MACS1149\_JD1 at $z=$9.11, both of which reveal intense [O{\sc iii}]88$\mu$m emission. In both cases we provide stringent upper limits on the presence of [C{\sc ii}]158$\mu$m with respect to [O{\sc iii}]88$\mu$m.  We review possible explanations for this apparent redshift-dependent [C{\sc ii}] deficit in the context of our recent hydrodynamical simulations. Our results highlight the importance of using several emission line diagnostics with ALMA to investigate the nature of the interstellar medium in early galaxies.
\end{abstract}

\begin{keywords}
galaxies : high-redshift, evolution, starburst - cosmology : early universe -  submillimeter: galaxies
\end{keywords}



\section{Introduction}

During the past few years the Atacama Large Millimetre/submillimeter Array (ALMA) has demonstrated its remarkable power by exploring the interstellar media (ISM) in galaxies in the reionisation era. In addition to studies of extreme and rare dusty sub-millimetre galaxies at redshifts $z\simeq$5-6 (e.g. \citealt{Capak2015}, \citealt{Pavesi2018}), the array has become the most reliable tool for spectroscopic confirmation of more typical distant star-forming galaxies (\citealt{Inoue2016}, \citealt{Laporte2017}, \citealt{Carniani2017}, \citealt{Smit2018}, \citealt{Hashimoto2018,Hashimoto2019}, \citealt{Tamura2018}).

The two most prominent emission features targeted by ALMA for normal star-forming galaxies are the  [O{\sc iii}]88$\mu$m and [C{\sc ii}]158$\mu$m fine structure lines, both of which are redshifted into the sub-mm atmospheric  window in the reionisation era. [C{\sc ii}]158$\mu$m is the dominant coolant of neutral gas in the ISM of local star-forming galaxies  and its luminosity is observed to correlate closely with star formation rate (SFR - \citealt{DeLooze2014}).  Early work exploring this relation at high-redshifts revealed increased scatter compared to that seen in  local samples. Whereas luminous Lyman break galaxies selected at $z\simeq$5-6 (e.g. \citealt{Capak2015}, \citealt{Willott2015}) as well as some Lyman-alpha emitters at $z\sim$6 (\citealt{Matthee2017}, \citealt{Carniani2018}, \citealt{Matthee2019}) found trends similar to those seen locally, other star-forming galaxies at $z>6$ often showed weak or no [C{\sc ii}]158$\mu$m detections (e.g. \citealt{Ota2014},  \citealt{Pentericci2016}). This so-called '[CII]-deficit' has been the subject of much debate and earlier discussed in the context of thermal saturation in ultra-luminous infrared galaxies \citep{Munoz2016}. While [C{\sc ii}]158$\mu$m is not affected by dust attenuation, it is sensitive to metallicity \citep{Olsen2017}, the ionisation state of the gas \citep{Vallini2017} and CMB attenuation. In addition, in a survey of three $z\simeq$7 sources, \cite{Maiolino2015}  discovered [C{\sc ii}]158$\mu$m emission with significant spatial offsets from the UV and Ly$\alpha$ emission, suggesting that the cores of young galaxies are disrupted by stellar feedback with line emission occurring only in external clumps of neutral gas. Although high-redshift data remains sparse and some non-detections are likely due to inadequate sensitivity, it remains of interest to pursue the topic to gain insight into the morphology and physical conditions in rapidly assembling young galaxies.

\begin{figure*}
\centering
\includegraphics[width=15cm]{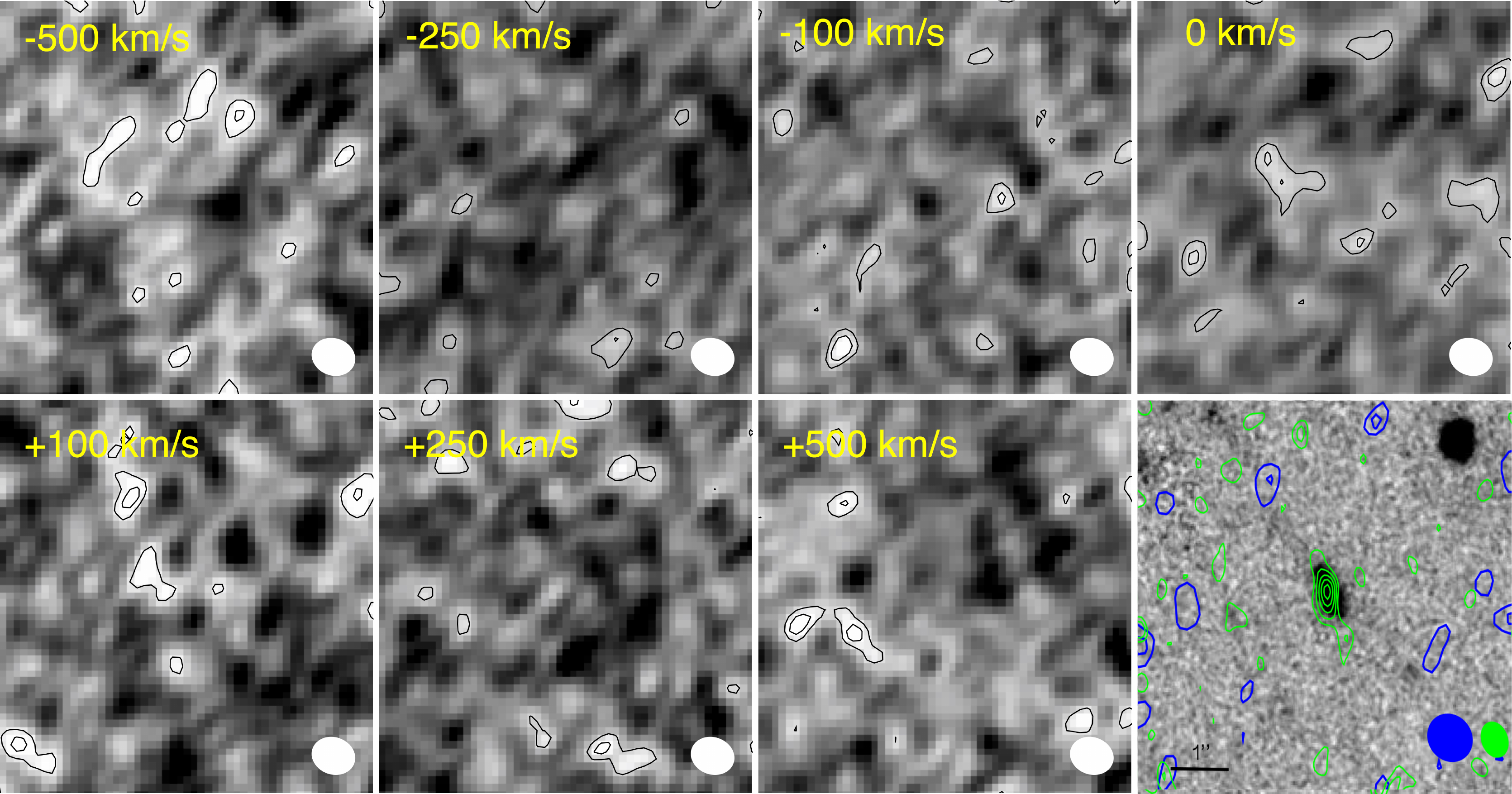}
\caption{\label{fig1} Search for [C{\sc ii}]158$\mu$m emission line near MACS1149\_JD1. Each stamp shows the flux contours (drawn from 2$\sigma$) at different velocity offsets (from -500km/s to +500km/s) with respect to the  [O{\sc iii}]88$\mu$m. redshift. The HST F160W image is shown at the bottom right of the figure with  [O{\sc iii}]88$\mu$m.  (green) and [C{\sc ii}]158$\mu$m (blue) contours. The shape of ALMA beam is placed at the bottom right of each ALMA stamp.   No [C{\sc ii}]158$\mu$m emission is evident.}
\end{figure*}

[O{\sc iii}]88$\mu$m emission also correlates with the star formation rate in local galaxies \citep{DeLooze2014} but, as a line with a higher ionisation potential, it is generated within H II regions rather than in photo-dissociation regions. The motivation for targeting [O{\sc iii}]88$\mu$m at high-redshift is  two-fold. \textit{Herschel} observations of dwarf galaxies suggested that it is a stronger line than  [C{\sc ii}]158$\mu$m in  low metallicity systems \citep{Madden2013}. Additionally, the line is well-placed observationally in  the ALMA bands at the very highest redshifts for which targets are available from deep \textit{Hubble} imaging. The line was prominently detected in two gravitationally-lensed targets, A2744\_YD4 at $z=8.38$ for which a dust continuum detection was also secured \citep{Laporte2017} and MACS1149\_JD1 at $z=9.11$ \citep{Hashimoto2018}. The two sources represent the highest redshift spectroscopically-confirmed sources accessible to ALMA and, in this paper we exploit the newly-available band 5 receiver to present new observations targeting [C II] 158$\mu$m in each source with the goal of further examining the relationship between [C{\sc ii}]158$\mu$m, [O{\sc iii}]88$\mu$m and various probes of star formation in early sources. Throughout the paper,  we adopt a $\Lambda$-dominated, flat Universe with  $\Omega_{\Lambda}$  = 0.7,  $\Omega_M$ = 0.3 and $H_0$ = 70 km s$^{-1}$ Mpc$^{-1}$.

\section{Observations}

Observations were carried out in band 5 during ALMA Cycles 5 and 6 under regular proposal (2017.1.00697 - PI: N. Laporte) and DDTs (2017.A.00026 and 2018.A.0004 - PI: N. Laporte). The lower spectral window used to observe A2744\_YD4 is centred on the frequency where  [C{\sc ii}]158$\mu$m is expected at $z=$8.38, and its width covers the redshift range 8.26 $< z <$ 8.43. The total exposure time on source was 3.8hrs.  A similar setup was used for the MACS1149\_JD1 observations, with a redshift range 8.96$\leq z \leq$ 9.16 and a total exposure time of 6.2hrs. Observations of A2744\_YD4 were made with the C43-2 configuration yielding a beam size of 1.3''$\times$0.79''. For MACS1149\_JD1, we used the configuration C43-4 to achieve a beam size of 0.75''$\times$0.63''. 
Data were reduced using the version 5.4.0 of the CASA pipeline (\citealt{CASA}), a Briggs weighting was applied in the \textit{tclean} task in both cases. For consistency purpose, we re-reduced ALMA band 7 data for A2744\_YD4 following the same procedures (2015.1.00594 - PI: N. Laporte)

\begin{figure}
\centering
\includegraphics[width=7cm, angle=0]{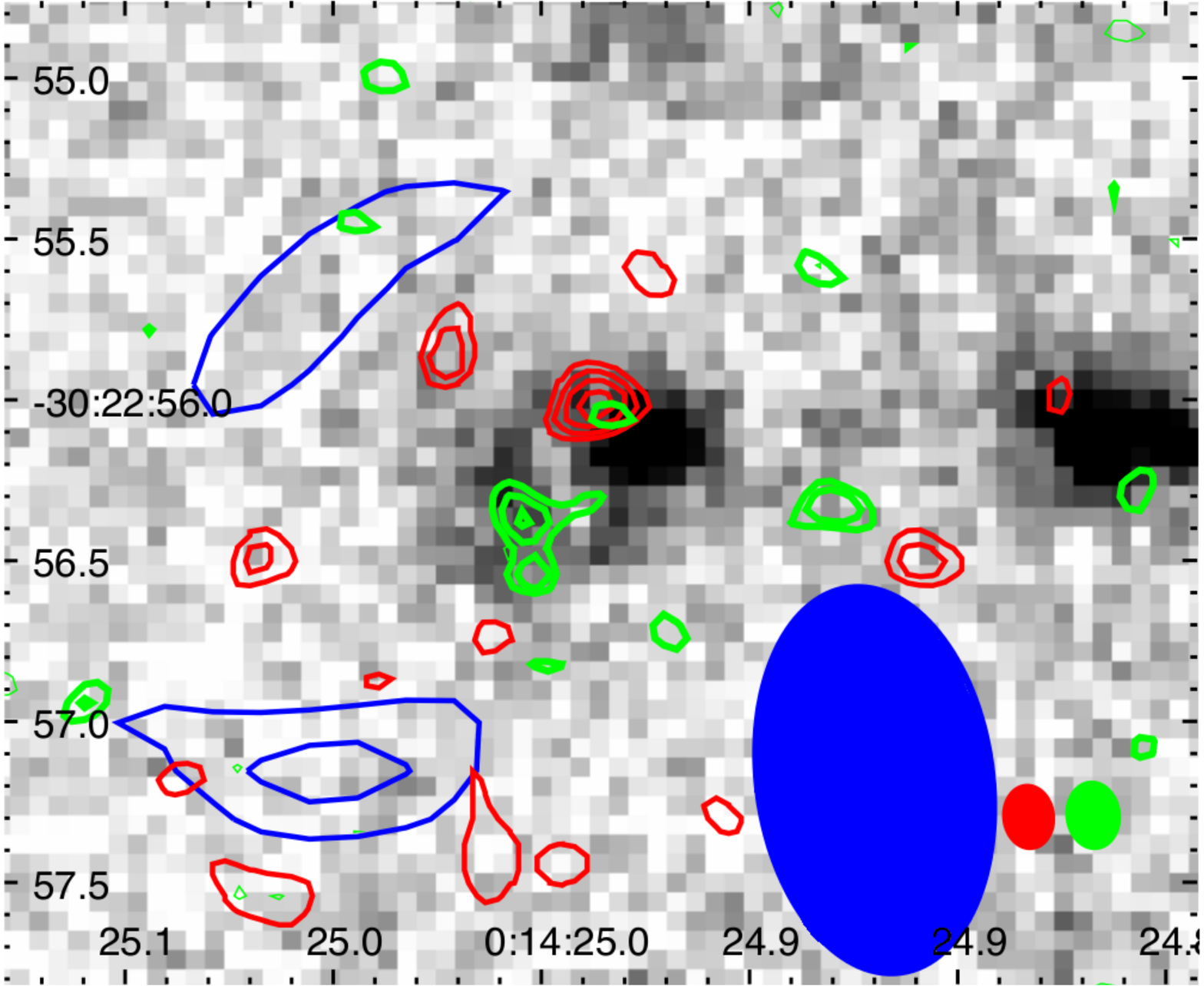}
\caption{\label{fig2} An ALMA view of A2744\_YD4 showing the respective positions of the dust detection in ALMA band 7 (red), the  [O{\sc iii}]88$\mu$m emission line (green) and the UV-rest frame continuum (HST/F160W image). The shape of each ALMA beam is shown at the bottom right. Contours are plotted from $2\sigma$. No  [C{\sc ii}]158$\mu$m emission (blue contours) is detected at more than $3\sigma$ near the rest-frame UV position of this galaxy. }
\end{figure}

We do not detect any band 5 continuum for either target. We measure 3$\sigma$ upper limits using several beam-size apertures distributed at the centre of the field where our targets are located, and find  $f_{\nu}^{158\mu m}$< 21 $\mu$Jy/beam  for A2744\_YD4 and $f_{\nu}^{158\mu m}$< 15 $\mu$Jy/beam for MACS1149\_JD1 (not corrected for magnification). We also searched for line emission in a 1.5'' radius circle around the UV-rest frame position of our targets (corresponding to a physical size of 13.2 and 14.1 kpc respectively for MACS1149\_JD1 and A2744\_YD4) and allowing a velocity offset respective to the  [O{\sc iii}]88$\mu$m redshift ranging from -500 km/s to 500km/s (e.g. \citealt{Hashimoto2019}). We rebinned the data assuming a FWHM of 100km/s for  [C{\sc ii}]158$\mu$m (as previously found for example in \citealt{Carniani2017}, \citealt{Smit2018}, \citealt{Bradac2017}). No emission is detected in either target (Figure~\ref{fig1} and Figure~\ref{fig2}) with a 3$\sigma$ upper limit on the  [C{\sc ii}]158$\mu$m luminosity of $L_{CII}^{JD1}$< 3.98$\times$10$^6$$\times$(10/$\mu$)L$_{\odot}$ and $L_{CII}^{YD4}$< 2.0$\times$10$^7$$\times$(2/$\mu$)L$_{\odot}$, assuming a FWHM=100km/s,  with the rms measured in several beam size apertures (with $\theta_{min}$=0.63'' and $\theta_{maj}$=0.75'' for JD1 and $\theta_{min}$=0.73'' and $\theta_{maj}$=1.21'') distributed in a 1.5'' radius circle around the UV restframe position and taking into account the best magnification for the two targets ($\mu$=2 and $\mu$=10 respectively for YD4 and JD1 - see \citealt{Laporte2017} and \citealt{Hashimoto2018} for details). We also applied the same method to more finely binned data (FWHM=50km/s) taking into account the FWHM of the [O{\sc iii}] 88$\mu$m line found in A2744\_YD4, but no emission line was found on either dataset. 

We summarise the salient properties of A2744\_YD4 and MACS1149\_JD1 in Table~\ref{tab1}. A similar non-detection of [C{\sc ii}]158$\mu$m was reported by \cite{Inoue2016} for a Lyman-$\alpha$ emitter at $z$ = 7.2 with [O{\sc iii}]88$\mu$m emission and, in the following analysis, we include those measurements.

\begin{table}
    \centering
    \begin{tabular}{l|cc}
            & A2744\_YD4    & MACS1149\_JD1 \\ \hline
$z_{OIII}$  & 8.382$^a$         & 9.1096$^b$ \\
L$_{OIII}$ ($\times$10$^7$L$_{\odot}$)   & 7.0$\pm$1.7$^a$ & 7.4$\pm$1.6$^b$ \\
L$_{FIR}$ ($\times$10$^{10}$L$_{\odot}$)  & 12.6$\pm$ 5.5$^a$  & $<$ 0.77$^b$ \\
L$_{CII}$ ($\times$10$^7$L$_{\odot}$)   & $<$ 2.0 (3$\sigma$)  & $<$ 0.4 (3$\sigma$) \\
$S_{\nu}^{158\mu m}$ ($\mu$Jy/beam)              & $<$ 10.5 (3$\sigma$)   & $<$1.5 (3$\sigma$) \\ 
$S_{\nu}^{88\mu m}$ ($\mu$Jy/beam)              & 99 $\pm$ 23 $^a$   & $<$ 5.3$^b$ (3$\sigma$) \\ 

\hline

SFR (M$_{\odot}$/yr)    &   20.4$^{+17.6}_{-9.5}$$^a$  &   4.2$^{+0.8}_{-1.1}$$^b$ \\
M$_{\star}$ (10$^9$M$_{\odot}$) &   2.0$^{+1.5}_{-0.7}$$^a$ &   1.1$^{+0.5}_{-0.2}$$^b$
    \end{tabular}
    \caption{  \label{tab1} Properties of the two $z>$8 galaxies reported in this paper. All values are corrected for magnification assuming $\mu$=2 for A2744\_YD4 and $\mu$=10 for MACS1149\_JD1. \\
    $^a$ \protect\cite{Laporte2017} \\ 
    $^b$ \protect\cite{Hashimoto2018} \\     
}
\end{table}

\section{Analysis}

\begin{figure}
\hspace{-1cm} \includegraphics[width=7.5cm, angle=-90.]{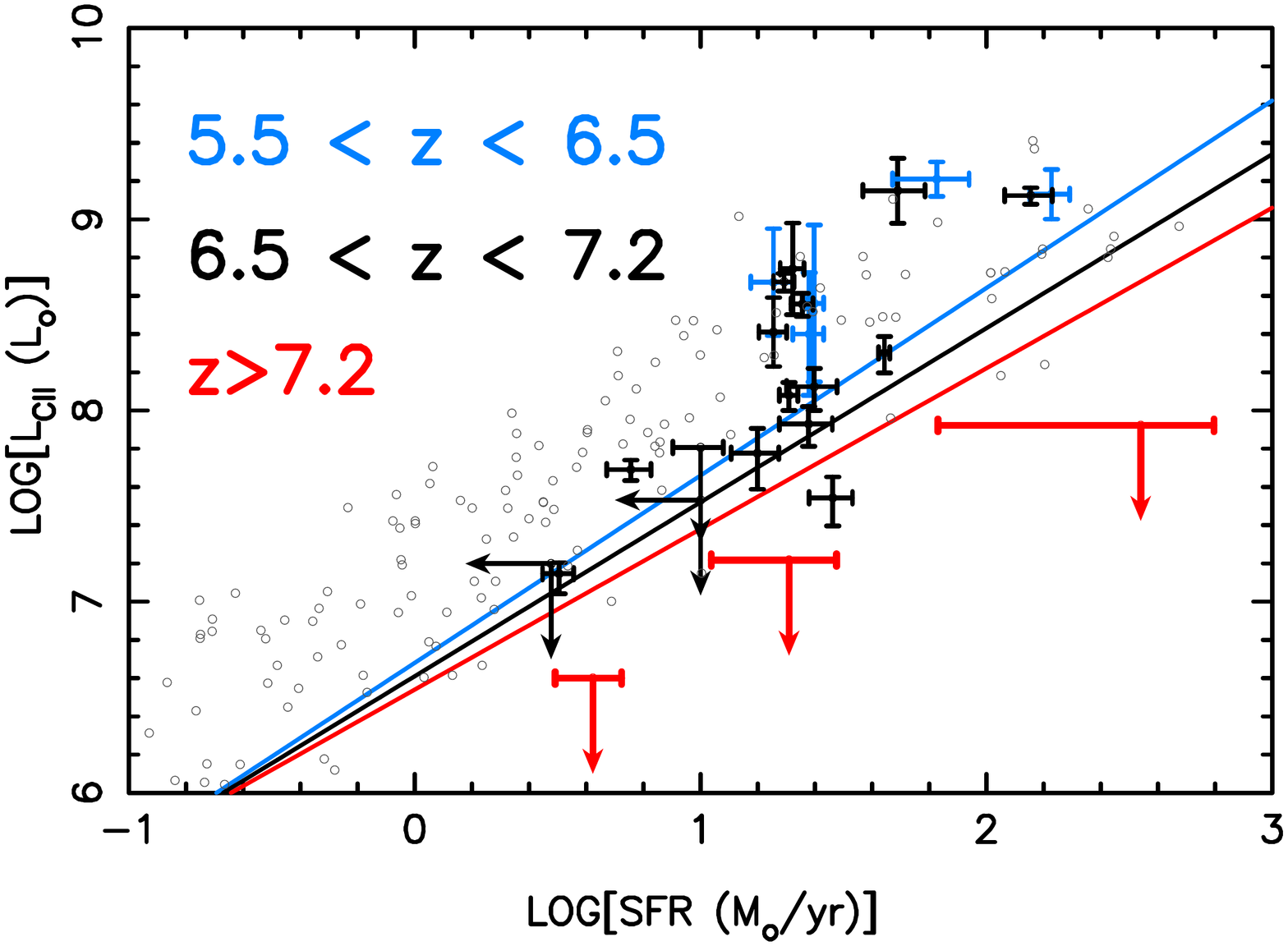}
\caption{\label{fig3} Relation between L$_{CII}$ and the SFR for the two galaxies studied in this letter plus that of \protect\citet{Inoue2016} (red) and previous $5.5<z<7.5$ galaxies studies from \protect\cite{Capak2015}, \protect\cite{Carniani2017}, \protect\cite{Carniani2018}, \protect\cite{Smit2018}, \protect\cite{Pentericci2016}, \protect\cite{Hashimoto2019},  \protect\cite{Kanekar2013}, \protect\cite{Ota2014}, \protect\cite{Bradac2017} and \protect\cite{Matthee2017} grouped according to redshift. Open circles show the location of local metal poor dwarfs galaxies \protect\citep{Madden2013}. We also plot the relation predicted by \protect\cite{Lagache2018} at $z\sim$6 (blue), 7 (black) and 8 (red).}
\end{figure}

\begin{figure}
 \hspace{-1cm} \includegraphics[width=8.0cm,angle=-90]{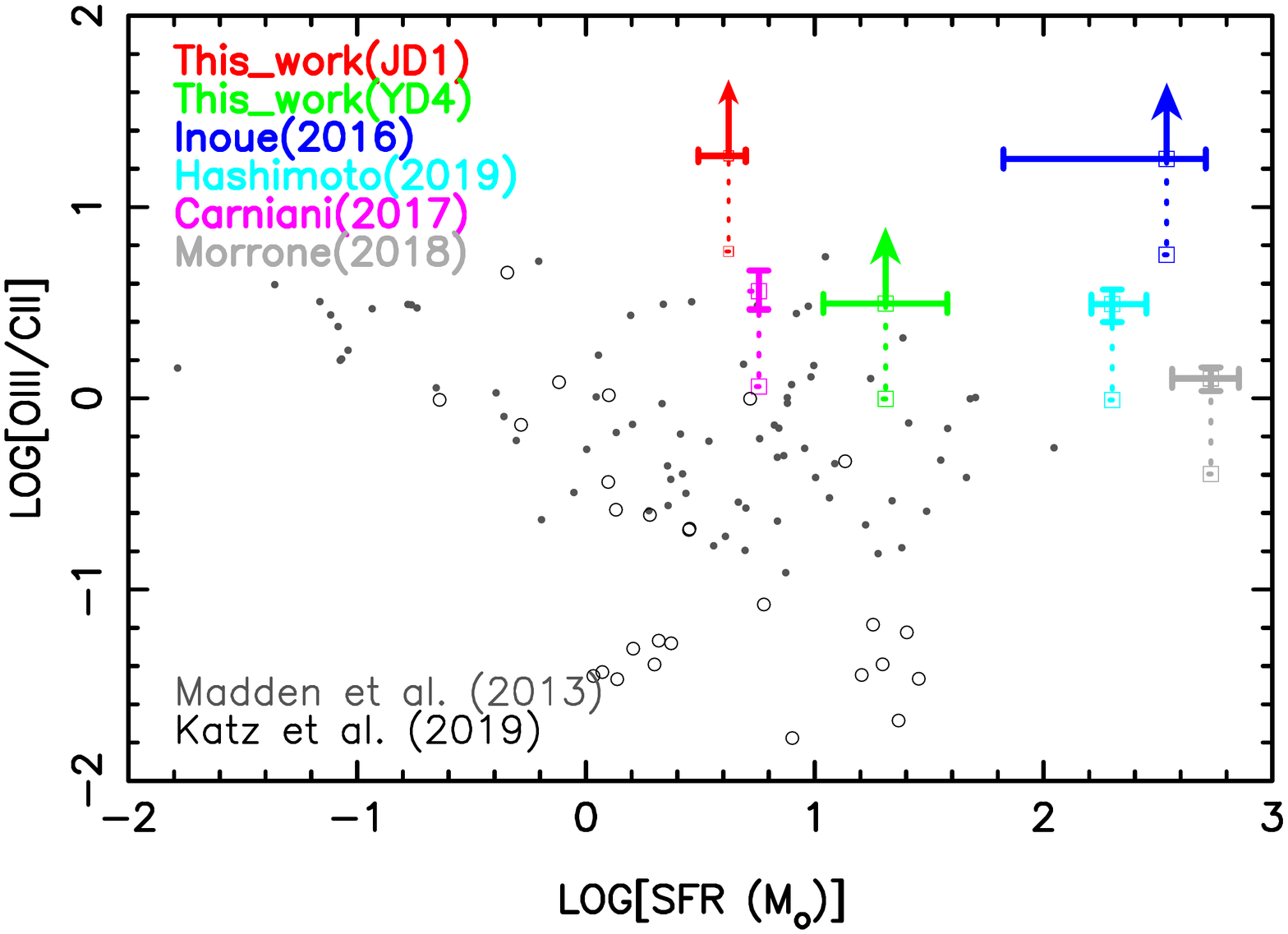}
\caption{\label{fig4} The [O III] / [C II] emission line ratio for high redshift galaxies. Our work on MACS1149\_JD1 and A2744\_YD4 together with the $z=7.2$ LAE \citep{Inoue2016} indicate ratios well above those seen in local metal poor dwarfs (\citealt{Madden2013}, grey circles) as well as numerical simulations capable of predicting both lines (\citealt{Katz2019}, black open symbols). The maximum effect of CMB attenuation is indicated by dashed lines below the current limits (see text for details).}
\end{figure}

In Figure~\ref{fig3} we compare the location of the two objects discussed in this paper, plus that of \cite{Inoue2016}, in the [CII] - SFR relation traced at lower redshift.  The apparent trend towards a [C II] deficit in the reionisation era is striking.  Likewise, Figure~\ref{fig4} shows the [O{\sc iii}] /[C{\sc ii}] line ratio in the context of lower redshift metal-poor dwarf galaxies \citep{Madden2013} and recent numerical simulations of high-redshift galaxies targeting both emission lines \citep{Katz2019}. The gas-phase metallicity in these simulations is 0.1 solar, comparable to that observed in the local dwarfs. Reducing the metallicity by a factor of 10 would be required to explain the absence of  [C{\sc ii}]158$\mu$m although at that point [O{\sc iii}]88$\mu$m emission would be similarly reduced. Although it is possible that the [C{\sc ii}]158$\mu$m and [O{\sc iii}]88$\mu$m emission regions are physically distinct in some of our sources, these comparisons suggest that a low metallicity may be insufficient to explain the deficit. Additionally, the strongest likely attenuation of [C{\sc ii}]158$\mu$m by cosmic microwave background radiation \citep{Lagache2018} seems unable to explain the size of the discrepancy (see dashed lines in Figure~\ref{fig4}). 

\begin{figure}
\hspace{-1.00cm} \includegraphics[width=10cm, angle=0]{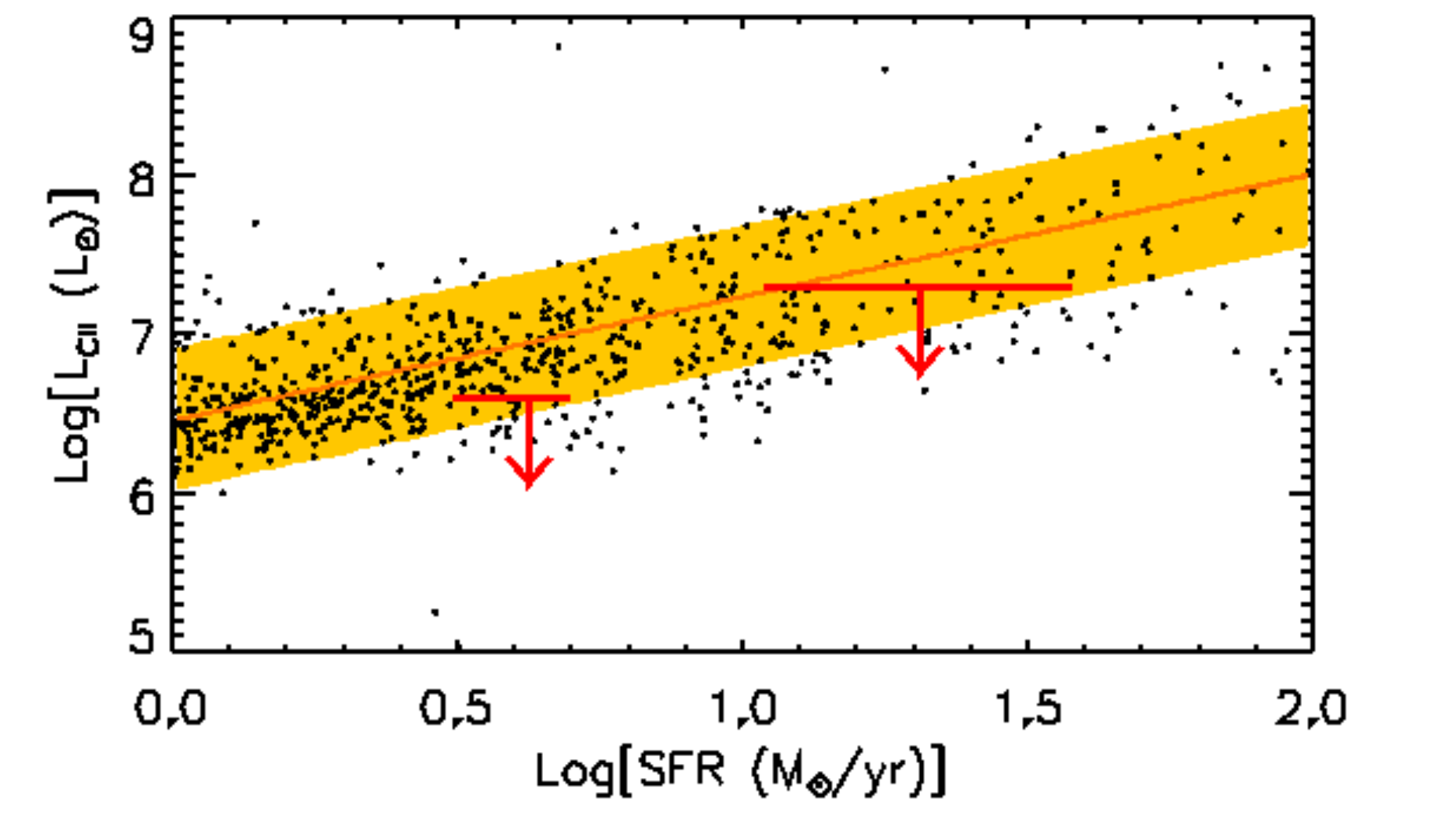}
\vspace{-0.50cm}
\caption{\label{fig5} As Fig \ref{fig3} with the location of the two $z\geq$8 galaxies discussed in this paper represented by red arrows. Black dots show the distribution of all of simulated galaxies from \protect\cite{Lagache2018}, extrapolated from their highest redshift  $z=$7.6 to redshift $z=$9 by estimating a mean CMB attenuation factor on the [C{\sc ii}]158$\mu$m luminosity from z=7.6 to z=9. The red line displays the mean relation between the SFR and L$_{CII}$ and the yellow region shows the mean dispersion (0.45 dex according to Fig.8 of \protect\citealt{Lagache2018}) of the simulated galaxies. Clearly both galaxies are extreme outliers in the relation.}
\end{figure}

Energetic feedback from intermittent star formation may be capable of expelling neutral gas and thereby suppressing [C{\sc ii}]158$\mu$m emission. Although the presence of a significant dust mass in A2744\_YD4 might then be considered surprising, the possibility of a spatial offset between [O{\sc iii}]88$\mu$m. emission and the dust continuum (Figure~\ref{fig2}) may imply regions with different physical conditions or represent the result of some feedback process.
One way to understand if a deficit of neutral gas is expected at high redshift is to determine the range of  [C{\sc ii}]158$\mu$m  emission expected in simulations. Examining a recent semi-analytical model of galaxy evolution \citep{Lagache2018} in over 10$^3$ simulated objects at $z\simeq$8 (Figure~\ref{fig5}) and focusing now only on the two highest-redshift sources, A2744\_YD4 and MACS1149\_JD1, we find  75 simulated objects that have extreme properties similar to A2744\_YD4 (i.e. SFR from  1 to 35 $M_{\odot}$ yr$^{-1}$; L$_{[CII]}<$2.0$\times$10$^7$L$_{\odot}$ ; $\log$(M$_{\star}$ [M$_{\odot}$]) from 8.8 to 9.7) with as mean properties <M$_{\star}$>=1.3$\times$10$^9$ M$_{\odot}$, <L$_{[CII]}$>=9.4$\times$10$^6$ and gas-phase metallicity <Z$_g$>=0.20.  Furthermore, only 8 simulated sources have [C II]158$\mu$m properties similar to MACS1149\_JD1 (i.e. SFR from 0.9 to 6.6 $M_{\odot}$ yr$^{-1}$ ; L$_{[CII]}<$0.4$\times$10$^7$L$_{\odot}$ ; $\log$(M$_{\star}$  [M$_{\odot}$]) from 8.7 to 9.4) with mean properties : <M$_{\star}$>=7.7$\times$10$^8$ M$_{\odot}$, <L$_{[CII]}$>=2.7$\times$10$^6$ L$_{\odot}$ and <Z$_g$>=0.25.  Since our observational upper limits are 3$\sigma$, this demonstrates the difficulty of reproducing our first glimpse at the weak  [C{\sc ii}]158$\mu$m  emission in $z>8$ sources.

A further explanation may be a trend towards higher ionisation parameters at early times \citep{Katz2016} for which there is some evidence in rest-frame UV spectroscopy of similar $z>7$ sources \citep{Mainali2018}. Such a trend may arise from a moderate non-thermal component or an increasing contribution from metal-poor massive stars. The original motivation for this study was to assemble of multi-line data using ALMA for sources in the reionisation era largely to test such hypotheses. Our discovery of a surprising  [C{\sc ii}]158$\mu$m  deficit argues for continuing this effort including further diagnostic lines sensitive to the nature of the radiation field, the gas-phase metallicity and the presence of neutral gas.

Finally, utilising the non-detection of the continuum of A2744\_YD4 in ALMA band 5 we have the opportunity to re-analyse the SED of this object. We include data from a previous ALMA band 6 programme covering the position of this target (2015.1.00463.S - PI : M. Ouchi). In this dataset, A2744\_YD4 is also not detected and we measured in a beam-size aperture a 2$\sigma$ upper limit flux of 30 $\mu$Jy/beam (not corrected for magnification). Using \textit{MAGPHYS} \citep{MAGPHYS}, we can give a first constraint on the dust temperature in this object $T_{dust}$ > 55 K. This value contrasts with the value generally used to determine the dust properties at high-$z$ (T$\sim$30K), but is consistent with recent simulations  (e.g. \citealt{Behrens2018}) which predict a higher dust temperature at high redshifts. Using the 3$\sigma$ upper limits for both band 5 and 6 observations decreases the minimum dust temperature to T$>$43 K.

\section{Summary}

The recent commissioning of the ALMA band 5 receiver has opened a new window to study the ISM of the two most distant gravitationally-lensed galaxies detected with ALMA band 7, namely A2744\_YD4 ($z=$8.38) and MACS1149\_JD1 ($z$=9.11). We have used this capability to search for the FIR emission line  [C{\sc ii}]158$\mu$m , the primary coolant of the ISM at low redshift, which should give valuable insight into the metallicity and neutral gas content for systems of known SFR. However, despite adequately sensitive data considering the [C{\sc ii}] - SFR relation observed at lower redshifts (e.g. $z<$6), neither of these targets is detected in the dust continuum or line emission.  Noting the magnification for these two targets ($\mu\sim$2 and 10 for A2744\_YD4 and MACS1149\_JD1 respectively), these non-detections imply [CII]158$\mu$m luminosities well below what is observed for $z\sim$0 metal poor dwarfs, reviving the discussion of a `[CII] deficit' previously considered at lower redshift. Likewise when studying the   [O{\sc iii}]88$\mu$m/  [C{\sc ii}]158$\mu$m  line ratio, we find anomalously high values. We examine this line ratio with a recent hydrodynamical simulation of the ISM in early galaxies \citep{Katz2019} and suggest that a low gas-phase metallicity may not be the sole explanation for this [C II] deficit. Other hypotheses include a high ionisation parameter consistent with trends seen in UV spectroscopy of similar $z>7$ sources or the suppression of neutral gas and hence [C{\sc ii}]158$\mu$m emission via energetic feedback from intermittent star formation. Using a semi-analytical model of galaxy evolution \citep{Lagache2018}, we demonstrate that such faint  [C{\sc ii}]158$\mu$m luminosities are  rarely expected at $z\geq$8. Further multi-line data on $z>8$ sources will be helpful in resolving this puzzle. \text\bf{Our study emphasises the importance of gathering multi-line ALMA data for sources in the reionisation era to robustly study the physical conditions in their interstellar media.}

\section*{Acknowledgements}

We thank Morgane Cousin to provide CMB attenuation estimates at $z\sim$9. NL and RSE acknowledge funding from the European Research Council (ERC) under the  European Union Horizon 2020 research and innovation programme (grant agreement No 669253). FEB acknowledges support from CONICYT-Chile (Basal AFB-170002, Programa de Cooperaci{\'{o}n} Cient{\'{\i}}fica ECOS-CONICYT C16U02, FONDO ALMA 31160033) and the Ministry of Economy, Development, and Tourism's Millennium Science Initiative through grant IC120009, awarded to The Millennium Institute of Astrophysics, MAS.
AKI and TH acknowledge funding from NAOJ ALMA Scientific Research Grant number 2016-01 A and JSPS KAKENHI Grant Number 17H01114. This paper makes use of the following ALMA data: ADS/JAO.ALMA\#2015.A.00463, ADS/JAO.ALMA\#2017.1.00697, ADS/JAO.ALMA\#2017.A.00026 and ADS/JAO.ALMA\#2018.A.0004,  ALMA is a partnership of ESO (representing its member states), NSF (USA) and NINS (Japan), together with NRC (Canada), MOST and ASIAA (Taiwan), and KASI (Republic of Korea), in cooperation with the Republic of Chile. The Joint ALMA Observatory is operated by ESO, AUI/NRAO and NAOJ.




\bibliographystyle{mnras}
\bibliography{mnras_template.bib} 



\bsp	
\label{lastpage}
\end{document}